\documentclass[preprint,5p,times,twocolumn]{elsarticle}

\usepackage{amsmath, amssymb,epsfig}
\usepackage{epstopdf}
\usepackage{color}

\newcommand{\be}{\begin{equation}}
\newcommand{\ee}{\end{equation}}
\newcommand{\bea}{\begin{eqnarray}}
\newcommand{\eea}{\end{eqnarray}}

\newcommand{\as}[1]{\overset{\text{a.s}}{#1}}
\newcommand{\Char}[2]{\chi_{#1}\left(#2\right)}

 \graphicspath{
{FIGURES/}
}

\journal{Physica D}

\begin{document}

\begin{frontmatter}

\title{Linearization effect in multifractal analysis:
Insights from the Random Energy Model}


\author[ENSL]{Florian Angeletti}
\author[LPTMS]{Marc M\'ezard}
\author[ENSL]{Eric Bertin}
\author[ENSL]{Patrice Abry}

\address[ENSL]{Universit\'e de Lyon, Laboratoire de Physique, ENS Lyon, CNRS,
46 All\'ee d'Italie, F-69007 Lyon, France, \\{\tt firstname.lastname@ens-lyon.fr}}

\address[LPTMS]{Laboratoire de Physique Th\'eorique et Mod\`eles Statistiques,
CNRS and Universit\'e Paris-Sud, B\^at.~100, F-91405 Orsay Cedex, France, {\tt mezard@lptms.u-psud.fr}.}

\begin{abstract}
The analysis of the linearization effect in multifractal analysis, and hence of the estimation of moments for multifractal processes, is revisited borrowing concepts from
the statistical physics of disordered systems, notably from the analysis of the so-called Random Energy Model.
Considering a standard multifractal process (compound Poisson motion), chosen as a simple representative example, 
we show:  i) the existence of a critical order $q^*$ beyond which moments,
though finite, cannot be estimated through empirical averages, irrespective of the sample size of the observation; 
ii) that multifractal exponents necessarily behave linearly in $q$, for $q > q^*$. 
Tayloring the analysis conducted for the Random Energy Model to that of Compound Poisson motion, we provide explicative and quantitative predictions for the values of $q^*$ and for the slope controlling the linear behavior of the multifractal exponents. 
These quantities are shown to be related only to the definition of the multifractal process and not to depend on the sample size of the observation.
Monte-Carlo simulations, conducted over a large number of large sample size realizations of compound Poisson motion, comfort and extend these analyses. 
\end{abstract}

\begin{keyword}
Multifractal analysis, linearization effect, compound Poisson motion, Random Energy Model, truncated moments, moment dominant contributions.
\end{keyword}

\end{frontmatter}


\section{Introduction}
\label{sec:intro}

Multifractal analysis is now considered as a canonical tool to study scaling properties and regularity fluctuations in time series (or n-dimensional fields) \cite{Riedi2003,j04,waj07}. 
Practically, it essentially amounts to computing time or space averages of
(the $q-$th power of) time and scale-dependent quantities $T(a,t)$,
leading to the so-called structure functions,
$S_n(a,q) = \frac{1}{n} \sum_{k=1}^{n} |T(a,t_k)|^q$.
The multiresolution quantities $T(a,t)$ are computed directly from the data,
and depend both on the time (or space) location and on the analysis scale $a$. 
Typical examples of such quantities $T(a,t)$ are the increments $X(t+a)-X(t)$
of a signal $X$ \cite{Frisch1995,fp85}, the wavelet coefficients \cite{arneodo1995}
or the wavelet Leaders \cite{waj07}.
In practice, multifractal analysis assumes that the structure functions behave
as power laws with respect to the analysis scale $a$, in a range $a_m < a < a_M$, with $a_M/a_m \gg 1$,
\begin{equation}
\label{eq-scaling}
S_n(a,q) \simeq S_0(q)\, a^{\zeta(q)},
\end{equation}
and to estimating the corresponding scaling exponent, $\zeta(q)$. 
The exponent $\zeta(q)$ is a concave function of the statistical order $q$.

It has been observed and argued that the exponent $\zeta(q)$ necessarily behaves as a linear function of $q$ beyond some value --see \cite{Molchan1996,Molchan1997} for the original reports of the phenomenon, \cite{Ossiander2000} for its analysis in the case of Mandelbrot multiplicative cascades, and \cite{lac04,ABRY:2007:B} (and \cite{BacryGlotterHoffmannMuzy2010} respectively) for more recent signal processing (and statistical analysis, respectively)  oriented contributions, in framework of the multifractal analysis of sample paths of stochastic processes.
Following \cite{lac04}, this is referred to as the \emph{linearization effect} in multifractal analysis,
and its study constitutes the core of this contribution, where it is intended to take advantage of a formal analogy between the linearization effect
in multifractal processes and the glass transition in the Random Energy Model (REM)
 \cite{Derrida} to interpret the linearization effect as a phase (or glass) transition.

The REM consists in a simple model classically used in statistical physics
as an illustration of a mean-field scenario for the glass transition in spin-glasses \cite{Derrida} or supercooled liquids \cite{Kirkpatrick1989,BouchaudBiroli}.
In this model, all microscopic configurations have
random independent energies $E_i$, drawn from the same distribution.
These energies are quenched, i.e., they do not evolve with time.
The interest of the REM stems from the fact that it displays a glass transition
at finite temperature, and that this transition can easily be studied analytically
\cite{Derrida}.
The physics underlying the glass transition in the REM
is rather simple. Above the glass transition temperature, thermal activation
is efficient and a large number of microscopic configurations are explored:
the system is in a 'liquid' state. Below the glass transition temperature,
thermal activation no longer plays a significant role, and the system
is frozen in the few lowest energy configurations. As a result, its entropy
per degree of freedom vanishes.
The definition and main properties of the REM are briefly recalled in \ref{app-REM}. 
The analogy between statistical physics models and multifractal analysis has been continuously and fruitfully used after the seminal contribution of Parisi and Frisch \cite{fp85} and the developments reported in \cite{arneodo1995}. 
The linearization effect has been studied in the light of statistical physics models such as those of \emph{progressive waves} \cite{MajumdarKrapvisky2000,CarpentierLeDoussal2001}, which can be regarded as alternative to the REM. 
This is notably the case in models such as \cite{MuzyBacryKozhemiak2006,MuzyBacryBailePoggi2008}.

The rationale underlying the comparison between multifractal analysis and REM
lies in two key facts: Both the REM and multifractal analysis involve the evaluation of sums of random variables
raised to a given power (constituting the control parameter of the problem); In both cases, these random variables have heavy-tailed distributions though all their moments are finite,
a typical example being the lognormal distribution.
In multifractal analysis, the structure functions are defined
as functions of the statistical order $q$, in the limit of a large number of terms.
In the REM, the partition function, $Z = \sum_k \exp(-\beta E_k)$, is defined
as a function of the inverse temperature $\beta$,
assuming that the number of microstates is large.
Hence the quantities $|T(a,t_k)|$ and $\exp(-E_k)$ formally play a similar role,
and the partition function $Z$ is the formal analog of the structure function $S_n(a,q)$.  
Quite importantly, heavy-tailed distributions have the property that the dominant terms
in the sum become very large (especially for large values of $q$ or $\beta$
in the present context), which
turns the use of the Central Limit Theorem and of the Law of Large Numbers
into a delicate matter (see, e.g., \cite{BenArous} in the context of the REM). 

The present contribution aims at exploring the extent
to which the statistical physics arguments, 
involved into the study of the REM to explain the zero-entropy phase transition, enable to
understand the linearization effect in multifractal analysis. 
This contribution thus further complements and enriches the connections between multifractal analysis and statistical physics \cite{fp85,PaladinVulpiani1987,arneodo1995,Frisch1995}.
More precisely, given the observation of a finite number $n$ of samples,
taken from a single realization of a multifractal process,
the goal of this paper is to analyze, using statistical physics techniques, the critical
statistical order $q^*$ up to which the empirical average $S_n(a,q)$
allows for a correct estimate of the ensemble average $\langle |T(a,t)|^q \rangle$.
The very example of multifractal processes consists in the celebrated Mandelbrot multiplicative cascades (see, e.g., \cite{m74,Frisch1995,Riedi2003} for reviews). 
However, in the present work, use will be made of Compound Poisson Cascades, recently introduced in \cite{baman02} (see also \cite{Chainais:2005:A}), because they benefit of statistical properties that are easier to handle practically and theoretically: their increments are stationary and characterized by a continuous scale invariance property, i.e., Eq. (\ref{eq-scaling}) above holds for a continuous range of scales $a \in [a_m,a_M]$.

\section{Compound Poisson cascades}
\label{sec-cpc}

\subsection{Definition of the processes}

Compound Poisson cascade (CPC) and compound Poisson motion (CPM) were recently introduced by Barral and Mandelbrot \cite{baman02} and are now considered as reference multifractal processes. 
The CPC $Q_r(t)$ corresponds to a product of positive, independent and
identically distributed random variables $W_i$, referred to as multipliers,
and associated to randomly located points $(t_i,r_i)$ on a rectangle
\begin{equation}
I_{r,L} = \left\{(t',r'):r\leq r'\leq 1,\, -\frac{1}{2}\leq t' \leq L+\frac{1}{2} \right\}.
\end{equation}
More precisely, the CPC is defined for $r>0$ through 
\begin{equation}
    \label{eq-cpcdef}
    Q_r(t)= B_r(t) \prod_{(t_i,r_i)\, \in \, {\cal C}_r(t)} W_i,
\end{equation}
where only multipliers associated with points belonging to the cone
\begin{equation}
\label{def-cone}
{\cal C}_r(t) = \left\{(t',r'): r\le r'\le 1,\, t-\frac{r'}{2} \le t' \le t+\frac{r'}{2} \right\}
\end{equation}
are taken into account.
The points ${(t_i,r_i)}$ are drawn from a Poisson process with intensity measure $dm(r,t)$
on the rectangle $I_{r,L}$. The parameter $B_r(t)$ is a normalizing
constant such that $\langle Q_r(t) \rangle=1$, where $\langle \dots \rangle$ denotes
the ensemble average (or expectation) of the process.

It has been shown that CPC satisfy the following key relation:
\begin{equation}
\langle Q_r(t)^q \rangle = \exp\left[-\varphi(q)\, m({\cal C}_r(t))\right], \, q > -1
\end{equation}
where $m({\cal C}_r(t)) = \int_{{\cal C}_r(t)} dm(r',t')$ corresponds to the measure of the cone
${\cal C}_r(t)$, and where $\varphi(q)$ is defined as
\begin{equation}
\varphi(q) = (1-\langle W^q \rangle)-q(1-\langle W \rangle).
\end{equation}
For the sake of simplicity, we only consider the case of smooth concave functions $\varphi(q)$,
a typical example of which being the lognormal case $\varphi(q)=cq(1-q)$, with a constant $c>0$.
The CPM, $X(t)$, is obtained by integrating the CPC, $Q_r(t)$, over time, and by taking the limit $r \to 0$:
\begin{equation}
    \label{eq-a}
X(t) = \lim_{r \rightarrow 0} \int_0^t Q_r(s)\, ds.
\end{equation}
This definition yields a well-defined process on condition that $\varphi(1^-) \geq -1$ \cite{baman02}.

\subsection{Scaling and multifractal properties}
\label{sec-cpcprop}

The increments $T(a,t)$ of the CPM $X(t)$, defined as
\begin{equation}
T(a,t)= X(t+a)-X(t),
\end{equation}
with $a>0$, are positive, due to the positivity
of $Q_r(t)$ (cf. Eq.~(\ref{eq-a})). 
If the intensity measure of the Poisson process has the factorized form $dm(r,t)=g(r)drdt$,
the increments $T(a,t)$ correspond to a stationary random process \cite{Chainais:2005:A},
meaning that all the statistical properties of $T(a,t)$ do not depend on time $t$.
Interestingly, it has also been shown that the moments $\langle T(a,t)^q \rangle$
are finite only for $-1<q<q_c$, where $q_c$ is given by \cite{Bacry2003}:
\begin{equation}
\label{eq-qcc}
q_c = \sup\, \{q\geq 1: q + \varphi(q) - 1\geq 0\}.
\end{equation}
One then expects that the probability $P(T(a,t)\geq x)$ behaves
asymptotically as $P(T(a,t)\geq x) \sim  x^{-q_c}$ when $ x \rightarrow +\infty$,
and hence that the variables $T(a,t)$ are heavy-tailed.

In addition, when
$g(r)dr=c(dr/r^2 + \delta_{\{1\}}(dr))$ (as proposed in \cite{Bacry2003}), 
where $\delta_{\{1\}}(dr)$ denotes a point mass at $r=1$, the infinite divisibility underlying the construction of $X(t)$ implies the following scaling properties
\begin{equation}
\label{eq-scalingA}
\langle T(a,t)^q \rangle = C_q\, a^{\lambda(q)},
\end{equation}
for $-1<q<q_c$, with $\lambda(q) = q + \varphi(q)$ \cite{Bacry2003,Chainais:2005:A}.
Note that (\ref{eq-scalingA}) is valid for all $a$ in the interval $0<a<L$.

The multifractal spectrum $D(h)$ consists of the Hausdorff dimension of the set of points $t$ on the real-line
that possess the same singularity (or H\"older) exponent $h$:
\begin{equation}
\label{holder}
T(a,t) \simeq c a^{h(t)}, \quad a \rightarrow 0.
\end{equation}
The function $D(h)$ hence provides a {\em global} description of the {\em local}
fluctuations of a sample path of $X(t)$. For a thorough introduction to multifractal analysis, the reader is referred to e.g., \cite{j04}.

From the results obtained in \cite{baman02}, it can be inferred that the
multifractal spectrum $D(h)$ of the CPM can be derived from the concave Legendre transform of $\lambda(q)$, 
\begin{equation}
\label{legendre-concave}
\lambda^\star (h) = \inf_{q} \{q h - \lambda(q)\},
\end{equation}
and can be expressed as:
\begin{equation}
\label{eq-mf} D(h)= \left\{\begin{array}{cl}
1+\lambda^\star(h), &  \makebox{ if } 1+\lambda^\star(h) \geq 0, \\
 -\infty, & \makebox{ otherwise.}
 \end{array}\right.
\end{equation}

Also, it is interesting to quantify the dependence structure of $T(a,t)$.
The two-time correlation function of $T(a,t)$ has been shown to take the following form
\cite{VEDEL:2010:A}:
\begin{equation}
\label{eq-cpcdep}
\langle T(a,t) T(a,t+s) \rangle =
\sigma^2 \left( |s+a|^{\lambda(2)} + |s-a|^{\lambda(2)} - 2 |s|^{\lambda(2)} \right),
\end{equation}
where $\sigma^2$ is the constant 
\begin{equation}
\sigma^2 = \frac{1}{\lambda(2)\, (\lambda(2)-1)}.
\end{equation}
This two-time correlation function can be recast into the following form
\begin{equation}
\label{eq-corr-time}
\langle T(a,t) T(a,t+s) \rangle = a^{\lambda(2)}\, f\left(\frac{s}{a}\right)
\end{equation}
with
\begin{equation}
f(u) = \sigma^2 \left( |u+1|^{\lambda(2)} + |u-1|^{\lambda(2)} - 2 |u|^{\lambda(2)} \right).
\end{equation}
Eq. (\ref{eq-corr-time}) shows that the variables $T(a,t)$ are correlated
over a typical time scale $a$. This result will prove useful in Sec.~\ref{seq-Critical_order}.
Let us however emphasize that the correlation time $a$ appearing in the two-time
correlation function of $T(a,t)$ is induced by the ``measurement'' process itself,
that is, the fact that $T(a,t)$ corresponds to the increment of the signal $X(t)$
on a scale $a$. The original signal $X(t)$ is scale invariant, and thus has no
characteristic time scale.

\subsection{Large deviation properties}

The statistics of the increments $T(a,t)$ has been characterized by the
moments $\langle T(a,t)^q\rangle$, given in Eq.~(\ref{eq-scalingA}). It is also interesting to
characterize this statistics through the probability density of $T(a,t)$.
For reasons that will appear clearer later, it is convenient to
consider the random variable $h_a(t)$ defined as
\be
h_a(t) = \frac{ \ln T(a,t) }  { \ln a }.
\ee 
Note that, from Eq.~(\ref{holder}) above, $h_a(t)$ corresponds, in the limit of fine scale
$a \rightarrow 0$, to the H\"older exponent $h(t)$.
The probability density function of $h_a(t)$, for a given $t$, is denoted as $p_a(h)$.
It does not depend on time $t$ due to the stationarity of the process $T(a,t)$.

We wish to show that $p_a(h)$ obeys a large deviation form in the limit
$a \rightarrow 0$, namely
\begin{equation}
\label{eq-large-dev}
p_a(h) \approx e^{-|\ln a|\, \psi(h)}.
\end{equation}
A common way to derive a large deviation form is the
G\"artner-Ellis theorem \cite{HugoTouchette2009,Gartner,Ellis}, which also allows the explicit
expression of $\psi(h)$ to be determined. We first define
\begin{equation}
\mu(q) = \lim_{a \rightarrow 0} \frac{1}{|\ln a|} \, \ln \left< e^{q|\ln a| h_a(t)} \right>.
\end{equation}
The function $\mu(q)$ can be computed from Eq.~(\ref{eq-scalingA}), yielding
\begin{equation}
\label{eq-mu-lambda}
\mu(q) = -\lambda(-q), \quad -q_c<q<1.
\end{equation}
From the properties of $\lambda(q)$, it can be inferred that $\mu(q)$ is
a smooth convex function.
Assuming the existence of the large deviation function $\psi(h)$ introduced
in Eq.~(\ref{eq-large-dev}),
\footnote{If $\mu(q)$ was finite for all real $q$, the G\"artner-Ellis theorem would imply the existence of $\psi$. In the present case, where $\mu$ is finite only for $-q_c<q<1$, we can strictly speaking only conjecture that $\psi$ exists.}
the G\"artner-Ellis theorem leads to the following expression
\begin{equation}
\label{eq-rate-function}
\psi(h) = \sup_q \{ qh - \mu(q)\}.
\end{equation}
The existence of the limit moment $q_c$ implies that the previous equation is valid for $h>h_c$ such that 
\begin{equation}
\label{eq-hc}
h_c=\mu'(-q_c) = \lambda'(q_c).
\end{equation}
Using Eq.~(\ref{eq-rate-function}) and the property $\lambda(q_c)=1$ resulting from
Eq.~(\ref{eq-qcc}), $h_c$ can be characterized by 
\begin{equation}
\label{eq-hc-char}
 \psi(h_c)=1-q_c h_c, 
\end{equation}
a property that we mention for later use.

Note that $\psi(h)$ is the convex Legendre transform of $\mu(q)$,
which is more common than the concave Legendre transform appearing in
Eq.~(\ref{legendre-concave}).
Using Eq.~(\ref{eq-mu-lambda}), the two Legendre transforms can be related
in the following way:
\begin{align}
\psi(h) &= \sup_q \{qh+\lambda(-q)\} \\
&= \sup_{q'} \{-q'h+\lambda(q')\} \\
&= -\sup_{q'} \{q'h-\lambda(q')\},
\end{align}
with $q'=-q$, leading to $\psi(h)=-\lambda^{\star}(h)$,
or equivalently, $\psi(h) = 1- D(h)$, as long as $D(h)\ge 0$.
Finally, we note that the large deviation behaviour of $p_a(h)$ for $a \rightarrow 0$
can be rewritten as
\begin{equation}
p_a (h) \approx a^{\psi(h)},
\end{equation}
which closely matches the so-called thermodynamical multifractal formalism used for practical
multifractal analysis, and relying on the heuristic assumption $p_a(h) \sim a^{1-D(h)}$ 
\cite{fp85,arneodo1995,Frisch1995}.

We can also observe that for $ h < h_c $, $\lambda^\star$ develops a linear branch 
\begin{equation}
 \lambda^\star(h) = h q_c -1.
\end{equation}
This expression may differ from the rate function $\psi$, but at least provides the convex hull
of $\psi$, which is consistent with the infinite nature of the moments of $T$ for $q \geq q_c$: Indeed, if $p_a(h) \approx a^{- h q_c}$, then $P(T(a,t)\geq x) \sim  x^{-q_c}$.

\section{Critical order for empirical moment estimation}
\label{statphys}

We now assume that a single observation of the process $X(t)$ is available,
via a finite number of sampled times with a sampling period $\delta t$.
From this observation, $n$ coefficients $\{ T(a,t_k), k=1, \ldots, n\}$ are computed,
with $t_k=(k-1)\delta t$
(to simplify the presentation, we assume that $n$ is independent of $a$, though this would not be
strictly true in practice). 
The structure function can be rewritten as:
\begin{equation}
 S_n(a,q) = \frac{1}{n}{\sum_{k=1}^{n} T(a,t_k)^q}
= \frac{1}{n}{\sum_{k=1}^{n} e^{-q|\ln a|\, h_a(t_k)}}.
\end{equation}
In this section, we introduce a critical order $q^*$, up to which the time average $S_n(a,q)$
estimates correctly the ensemble average $\langle T(a,t)^q \rangle$, and we study how
$q^*$ behaves as $n \rightarrow +\infty $.
The reasoning relies on combining an estimate of the number of independent coefficients,
with two arguments inspired from the analysis of the REM
(see \cite{AngelettiBertinAbry2011} and \ref{app-REM}),
namely the identification of a dominant contribution
from a saddle-point estimation of the $q$-th moment, and
a truncation effect due to finite sample size observations.

\subsection{Dominant moment contribution}

The expression Eq.~(\ref{eq-scalingA}) of the moments of $T(a,t)$ can be easily recovered
from the large deviation form Eq.~(\ref{eq-large-dev}):
\begin{equation}
\label{eq-FctGen}
\langle {T(a,t)^q} \rangle =  \langle a^{q h_a(t)} \rangle
\approx \int_{-\infty}^{+\infty} e^{-|\ln a| [ qh+\psi(h) ]}\, dh.
\end{equation}
In the limit $a\rightarrow 0$, a saddle-point evaluation shows that the dominant
contribution to this integral is located at $h=h_{\mathrm{m}}$ given by
\begin{equation}
\label{eq-hstar}
\psi'(h_{\mathrm{m}})=-q,
\end{equation}
so that the moment $\langle {T(a,t)^q}\rangle$ reads
\begin{equation}
\label{moment-large-dev}
\langle {T(a,t)^q} \rangle \approx e^{-|\ln a|\, [ qh_{\mathrm{m}}+\psi(h_{\mathrm{m}}) ]}.
\end{equation}
For $h>h_c$,  $\psi(h)=-\lambda^{\star}(h)$ and $h_{\mathrm{m}}$ satisfies: 
\begin{equation}
\label{eq-hstar-lambda}
(\lambda^{\star})'(h_{\mathrm{m}})=q.
\end{equation}
From the properties of the Legendre transform, it also implies: 
\begin{equation}
\label{eq-psi-lambda-hm}
qh_{\mathrm{m}}+\lambda^{\star}(h_{\mathrm{m}}) = \lambda(q)
\end{equation}
and
\begin{equation}
\label{eq-hcirc}
h_{\mathrm{m}}=\lambda'(q),
\end{equation}
which implicitly defines $h_{\mathrm{m}}$ as a function of $q$.

\subsection{Finite sample size} 
\label{sec-finite_sample_size}

In the set $\{ h_a(t_k),\, k =1,\ldots, n \}$, the largest individual contribution
to $S_n(a,q)$ comes, when $a\rightarrow 0$, from the lowest value of $h_a$.
To quantify the order of magnitude of the typical lowest available value of this set, 
a simple idea is to consider a threshold $h^{\dag}_a(n)$ such that 
\begin{equation} \label{def-hbox}
P( h_a(t_1),\dots,h_a(t_n) > h^{\dag}_a) = e^{-\tau},
\end{equation}
where $\tau >0$ is an arbitrary constant.
The notation $P( h_a(t_1),\dots,h_a(t_n) > h^{\dag}_a)$ denotes the probability
that all the random variables $h_a(t_1),\dots,h_a(t_n)$ are larger than the value $h^{\dag}_a$.
If the random variables $\{h_a(t_k)\}$ were independent,
Eq.~(\ref{def-hbox}) would, for large $n$, simplify to: 
\begin{equation}
\label{hbox-iid}
P(h_a<h^{\dag}_a) \approx \frac{\tau}{n}.
\end{equation}
For the CPC, the variables $\{h_a(t_k)\}$ are strongly dependent.
We can however postulate that there exists an \emph{effective} number $n_a<n$
of independent samples.
We can then define $h^{\dag}_a$ by analogy to Eq.~(\ref{hbox-iid}), leading to
\begin{equation} \label{eq-hdag-na}
P(h_a<h^{\dag}_a) \approx \frac{\tau}{n_a},
\end{equation}
or equivalently,
\begin{equation}
\label{eq-hdag}
\ln \left( \frac{n_a}{\tau}\right) = - \ln P(h_a<h^{\dag}_a).
\end{equation}
Let us now determine $h^{\dag}_a$ more explicitly as a function of $n_a$.
Using the large deviation form
Eq.~(\ref{eq-large-dev}) in Eq.~(\ref{eq-hdag}), ones gets:
\begin{equation}
\ln \left( \frac{n_a}{\tau}\right) =  - \ln \left( \int^{h^{\dag}_a}_{-\infty} e^{-|\ln a|\, \psi(h)} dh \right).
\end{equation}
Because $\psi$ is a decreasing function of $h$ on this interval, a saddle-point argument amounts to evaluating the integral as the integrand boundary value:  
\begin{equation}
\label{eq-hdaga}
\ln \left( \frac{n_a}{\tau} \right) \simeq |\ln a| \,\psi ( h^{\dag}_a ).
\end{equation}
The threshold $h^{\dag}_a$ is thus determined from the implicit equation
\begin{equation}
\label{eq-hsquare0}
 \psi(h^{\dag}_a) =  \frac{1}{|\ln a|}\, \ln \left( \frac{n_a}{\tau} \right).
\end{equation}
Note that the arbitrary choice of  $\tau$ is fading away in the limit $a\rightarrow0$.

\subsection{Truncated moments and structure function} 
\label{sec-truncation}

Having introduced the threshold $h^{\dag}_a$,  truncated moments can be defined as: 
\begin{equation}
M(a,q) = \int^{+\infty}_{h^{\dag}_a}  a^{q h} p_a(h)\, dh.  
\end{equation}
Let us emphasize that the truncated moment in principle depends on the specific choice made for
the threshold $h^{\dag}_a$. This slight dependence however has no consequence on the conclusions
drawn from the truncated moments $M(a,q)$, as seen below.

These truncated moments provide us with a relevant evaluation of (the log of) the expectation of the random variables constituted by the structure functions. More precisely, if we analyse scale by scale the signal $X(t)$ by considering a sequence of scale $a_k=2^{-k} L$ with $n_k=2^k$ then 
\begin{equation}
\label{eq-as-conv-lnS} 
 \lim_{k\rightarrow +\infty} \frac{\ln S_{n_k}(a_k,q)} {|\ln a_k|}  \as = \lim_{a\rightarrow 0} \frac{\ln  M(a,q)}{|\ln a|}. 
\end{equation}

A proof of this result, which mainly relies on the Borel-Cantelli lemma under a realistic assumption,
is provided in \ref{app-as-conv}.
In a standard framework of i.i.d.~random variables, this limit would be far too rough to provide any useful insight about $S_n(a,q)$. In our case, however the limit retains some fundamental information about the behaviour of $S_n$.

\subsection {Critical order} 
 \label{seq-Critical_order}
Combining the truncation and saddle-point arguments, we
observe that two different situations can arise (cf. \cite{AngelettiBertinAbry2011}).

When $ h_{\mathrm{m}}(q) > h^{\dag}_a(n_a)$, the truncated moment $M(a,q)$,
and hence the structure function $S_{n}(a,q)$,
correctly accounts for the ensemble average $\langle T(a,t)^q \rangle$,
which can thus be evaluated as (using a saddle-point evaluation in the limit $ a \rightarrow 0$):
\begin{equation}
\label{eq-momHbis}
\lim_{a\rightarrow 0} \frac{\ln  M(a,q)}{|\ln a|} = -(qh_{\mathrm{m}}(q)+ \psi(h_{\mathrm{m}}(q)).
\end{equation}

In contrast, when $h_{\mathrm{m}}(q) < h^{\dag}_a(n_a)$, the dominant contribution
to the truncated moment is no longer
located at $h_{\mathrm{m}}(q)$ but instead comes from the lower bound $h^{\dag}_a(n_a)$
of the integration interval, in which case $\ln M(a,q)$
reads (again from a saddle-point evaluation when $ a \rightarrow 0$): 
\begin{equation}
\label{eq-momHter}
\lim_{a\rightarrow 0} \frac{\ln  M(a,q)}{|\ln a|}  = - (q h^{\dag}_a(n_a) + \psi(h^{\dag}_a(n_a)).
\end{equation}
This reveals that, for small enough $a$, $S_n(a,q)$ undergoes
a ``phase transition'' when varying $q$, occurring at a critical order $q^*_a $, defined as
\begin{equation}
\label{eq-hsquarehcirc}
h_{\mathrm{m}}(q^*_a) = h^{\dag}_a(n_a). 
\end{equation}
Interestingly, Eq.~({\ref{eq-momHter}) reveals a linear behaviour in $q$
of $ \ln S_{n}(a,q) $ when $q \geq q^*_a$, thus accounting for the linearization
effect reported in \cite{Molchan1996,Ossiander2000,lac04,ABRY:2007:B}.
Note also that Eq.~(\ref{eq-hsquarehcirc}) can alternatively be interpreted
as the minimal number of independent
samples $n^*_{a}(q)$ needed to correctly estimate the moment of order $q$.

We now investigate the asymptotic behaviour of $q^*_a$ in the limit
$n_a \rightarrow \infty$ (or $a \rightarrow 0$).
In practice, this limit is obtained by successively considering smaller and smaller resolutions
$\delta t$.
We have seen in Eq.~(\ref{eq-corr-time}) that the variables $T(a,t)$ are correlated
over a time $a$, so that the same result is expected for $h_a(t)$.
A natural estimate for $n_a$ is thus
\begin{equation}
\label{eq-na}
n_a = \frac{L}{a},
\end{equation}
where $L$ is the total length of the signal.

We first observe that Eqs.~(\ref{eq-hsquare0}) and (\ref{eq-na}) implies 
\begin{equation}
\label{eq-hsquare}
\psi(h^{\dag}_a) \rightarrow 1
\end{equation}
when $a \rightarrow 0$. Hence, $h^{\dag}_a(n)$ converges in the limit $a \rightarrow 0$
to a finite value $h^{\dag}_0$, independent of $n$, and uniquely determined by $\psi(h^{\dag}_0) = 1$
which, in the multifractal settings, can be rewritten as:
\begin{equation}
\label{eq-hsdagmf}
 D(h^{\dag}_0)=0.
\end{equation}
This result is particularly interesting from the point of view of multifractal analysis, and its interpretation will be further discussed in Section~\ref{sec-comments}

Eqs.~(\ref{eq-psi-lambda-hm}) and (\ref{eq-hcirc}) implicitly rely on the
assumption $h^{\dag}_0>h_c$, that we now briefly discuss.
Using Eq.~(\ref{eq-hc-char}) and the fact that $\psi$ is a decreasing function
from $h_c$ to $h^{\dag}_a$, one can see that the condition $h^{\dag}_0>h_c$ is equivalent to
\begin{equation*}
\label{eq-hc<hdag}
1 < 1 -h_c q_c .
\end{equation*}
Hence the property $\lambda'(q_c)=h_c<0$ implies $h^{\dag}_0>h_c$, a condition which is thus
true in all interesting cases, confirming the validity of Eqs.~(\ref{eq-psi-lambda-hm})
and (\ref{eq-hcirc}).
Combining these equations with Eqs.~(\ref{eq-hdaga}) and (\ref{eq-hsquarehcirc})
yields the relation:
\begin{equation}
\label{eq-critter}
\frac {\ln (n_a/\tau) } {\ln a}  = q^*_a \lambda'(q^*_a) -  \lambda(q^*_a ).
\end{equation}
Using $n_a=L/a$, we find
\begin{equation}
\label{eq-critterbis}
\frac {\ln (L/\tau) } {\ln a} -1   =  q^*_a \lambda'(q^*_a)- \lambda(q^*_a).
\end{equation}
In the limit $ a \rightarrow 0$, Eq.~(\ref{eq-critterbis}) defines a finite asymptotic
critical order $q^*$ as:
\begin{equation}
\label{eq-qc}
0  = 1 + q^* \lambda'(q^*) - \lambda(q^*).
\end{equation}

The comparison of Eq.~(\ref{eq-hsdagmf}) and Eq.~(\ref{eq-qc}) moreover immediately shows that: 
\begin{equation}
\label{eq-qstarhadga}
h^{\dag}_0  = \lambda'(q^*).
\end{equation}

In summary, assimilating $S_n(a,q)$ with $M(a,q)$ and
combining Eqs.~(\ref{eq-momHbis}) and (\ref{eq-momHter}), we find that the empirical
structure function
$S_n(a,q)$ typically behaves as a power law with respect to the analysis scale $a$
when $a \rightarrow 0$, namely $S_n(a,q) \sim S_0(q)\, a^{\zeta_{\mathrm{e}}(q)}$,
with $\zeta_{\mathrm{e}}(q)$ an empirical scaling exponent.
More formally, we can define the scaling exponent $\zeta_{\mathrm{e}}(q)$ as the random variable: 
\begin{equation}
\label{eqzetaq}
\zeta_{\mathrm{e}}(q) = \lim_{a \rightarrow 0}  \frac{\ln S_n(a,q)}{\ln a }.
\end{equation}
Eq.~(\ref{eq-as-conv-lnS}) then implies that $\zeta_{\mathrm{e}}(q)$ is almost surely equal
to its average value $\langle \zeta_{\mathrm{e}}(q) \rangle \equiv \zeta(q)$:
\begin{equation}
\label{eq-zeta-as}
\zeta_{\mathrm{e}}(q) \as{=} \zeta(q).
\end{equation}
The value of $\zeta(q)$ can be expressed, using Eqs.~(\ref{eq-momHbis}) and (\ref{eq-momHter}), as
\begin{equation}
\label{eqzetaqb}
\left.
\begin{array}{lcll}
\zeta(q) & = & \lambda(q), & -1 < q \leq q^*, \\
& & 1 + q \lambda'(q^*), & q > q^*.
\end{array}
\right\}
\end{equation}
From a practical viewpoint, the above results can be summarized as follows in terms of the structure function $S_n(a,q)$.
If $q<q^*$, 
\begin{equation}
\label{eq-lsnaqa}
	\ln S_n(a,q)\approx \lambda(q) \, \ln a ,
\end{equation}
while if $q>q^*$, 
\begin{equation}
\label{eq-lsnaqb}
	\ln S_n(a,q) \approx  (1+ q \lambda'(q^*) )  \, \ln a .
\end{equation}

\subsection{Comments on the critical order}
\label{sec-comments}

To sum up, Eqs.~(\ref{eq-qc}),  (\ref{eq-qstarhadga}),  (\ref{eqzetaqb}),  (\ref{eq-lsnaqa})  and  (\ref{eq-lsnaqb}) constitute the most important results of the present contribution,  
that call for a number of comments.

\indent i) For multifractal processes such as CPM,
the time averages (or structure functions) $S_n(a,q)$ do not converge at large $n$
to the ensemble average $\langle |T(a,t)|^q \rangle$, for $ q > q^*$.

\indent ii) It is important to note that $q^* \neq q_c$.
Using Eqs.~(\ref{eq-qc}) and (\ref{eq-qcc}), it can easily be shown that $q^* < q_c$. 
Therefore, the critical order up to which $S_n(a,q)$ accurately estimates the ensemble average is not related to the finiteness of the moments of $|T(a,t)|$ but occurs for much lower values
of $q$.

\indent iii) The critical order $q^*$ and the critical H\"older exponent $h^{\dag}_0$ are found to be independent of the actual number $n$ of available samples. 
Therefore, increasing $n$ (through a decrease of the sampling period $\delta t$) does not allow for a significantly better result.
Moreover, Eq.~(\ref{eq-critterbis}) shows that in practice the effective critical order at scale $a$ only weakly varies with $n$ or $a$.
Note that for given specific models $\lambda(q) $, it is possible that the solution of Eq. (\ref{eq-qc}) is $q^* = +\infty$, which can either be understood  as the fact that the linearization effect does not occur for such cases or (our preferred interpretation) that the linearization effect is a general effect that is rejected at infinity for those particular cases. 

\indent iv) The above properties, which appear as consequences of Eq.~(\ref{eq-hsdagmf}),
can be interpreted as follows, in a way closely paralleling the arguments
of the REM (see \ref{app-REM}). 
In a given sample, the number of independent points having a singularity exponent $h$ scales as $n_a\,e^{-\psi(h)|\ln a|}$, for $a \rightarrow 0$. 
Using $n_a = L/a$, the above number thus scales as
$e^{(1-\psi(h))\, |\ln a|}$. 
This means that in a given sample, there will be a large number of points with singularity $h$ when $1-\psi(h)>0$ (corresponding to $D(h)>0$), 
while there will be no such points in a typical sample when $1-\psi(h)<0$, irrespective of its observation duration $L$ and of the analysis scale $a$. 
The value $h_0^{\dag}$, such $D(h_0^{\dag})=0$ (cf.
Eq.~(\ref{eq-hsdagmf})), therefore receives a simple interpretation within this framework.
The analogy with the REM can even be pushed further.
Given the correspondance between, on one-hand, the partition function $Z$
of the REM and the structure function $S_n$ in the multifractal case,
and, on other-hand, the inverse temperature $\beta$ and the order $q$ of moments, 
we find that the analog of the entropy per degree of freedom (which is zero in the
low temperature phase of the REM $\beta>\beta_{\mathrm{g}} \equiv T_{\mathrm{g}}^{-1}$)
is the quantity $q\zeta'(q)-\zeta(q)+1$,
which is indeed equal to zero in the linear regime obtained for $q>q^*$.
Table~\ref{tabcomp} sketches the correspondence between the quantities defined in the REM and
in multifractal analysis
(see \ref{app-REM} for the definition of the notations used in the REM).\\
\indent v) It is worth mentioning that the interpretations of the analyses reported above in terms of $ h $ compared to $ h_0^{\dag} $ and entropy had already been envisaged by B. Mandelbrot in a series of seminal articles dedicated to detailed practical aspects of multifractal analysis, the most prominent of them being e.g., \cite{Mandelbrot1990, Mandelbrot2003}. 

\begin{table}
\begin{center}
\begin{tabular}{c|c}
\hline
\hline
 MF & REM \\
\hline
\hline
$S_n(a,q) \approx \frac{1}{n_a}\sum_{j=1}^{n_a} T(a,t_{k_j})^q$ & $ Z/2^N = \frac{1}{2^N}\sum_{j=1}^{2^N} e^{-\beta E_j}$ \\ \hline
$ \ln n_a \sim -\ln a $ & $ \ln n = N \ln 2$ \\ \hline
$q$ & $\beta=T^{-1}$ \\ \hline
$\ln T(a,t_{k_j}) $  & $-E_j $ \\ \hline
$  h_a(t_{k_j})$  & $\epsilon_j / \ln 2 $ \\ \hline
$ -q^{-1} \ln \big(n_a\, S_n(a,q)\big)$ & $F=-T\ln Z$ \\ \hline
$q \zeta'(q)-\zeta(q)+1$ & $S/(N\ln 2)$ \\ \hline
$q ^*$ & $\beta_g$ \\ \hline
$ h^{\dag}_0$ & $ \epsilon^{\dag} / \ln 2  $ \\ \hline
$h_m$ & $ \epsilon_m / \ln 2 $ \\ \hline
 \hline
 \end{tabular}
 \caption{\label{tabcomp} {\bf MF vs.~REM}. Mapping between quantities defined in the multifractal analysis (MF) and in the Random Energy Model (REM), valid in the limit of small $a$ and large $N$. In order to interpret $S_n(a,q)$ as a sum of (almost) independent variables, a sequence of $n_a$ times $\{t_{k_j}, j=1,\ldots,n_a\}$ is extracted from the full set $\{t_k, k=1,\ldots,n\}$.}
\end{center}
\end{table}

\indent vi) The theoretical analysis of the linearization effect obtained in the present contribution from REM-type statistical physics arguments is similar to (and hence fully conforts) the conjecture formulated in \cite{lac04,ABRY:2007:B},
stemming from the interpretation in terms of extreme values and local regularity of empirical observations obtained from the application of the multifractal formalism to numerical simulations of CPM and other related multifractal processes.

\indent vii) Complementary theoretical analysis (as in Section~\ref{statphys}) and numerical analysis (as in Section~\ref{sec-mc} below) conducted for multifractal processes other than CPM (not reported here) suggest that the results obtained here for CPM are valid for much broader classes of multifractal processes (cf., e.g., \cite{lac04}).

\section{Monte-Carlo simulations}
\label{sec-mc}

In the study of systems such as the REM, or, more generally, in the frozen phase of spin glasses, the condensation of explored configurations onto a small subset is classically measured using a theoretical or numerical tool referred to as the \emph{participation ratio} \cite{mezard1984,Mezard:SpinGlass}. 
In this section, we make use of this tool to further analyze the \emph{linearization effect} in the context of multifractal analysis. 

The definition of the participation ratio $\rho(a,q,p) $ is taylored from classical formulations in statistical physics to the context of multifractal processes, with $n_a = L/a$: 
\begin{equation}
\label{eq-partratio}
\rho(a,q,p) = \frac{\sum_{k=1}^{n_a} |T(a,ka)|^{qp}}{\left( \sum_{k=1}^{n_a} |T(a,ka)|^{q}\right)^p}.
\end{equation}

In the analysis of the REM, it can be shown that, in the glassy phase $\beta>\beta_g$
(associated here to $q > q^*$), the participation ratio is \emph{non-self-averaging},
which means that it depends on the explicit observation (or sample) of the process $X(t)$,
and hence of its increments $T(a,t)$, even in the limit $a \rightarrow 0$. 
Therefore, in that limit, its expectation satisfies the following explicit closed-form formula,
for all $p > 1$ and $q >0$, \cite{mezard1984,Mezard:SpinGlass}:
\begin{equation}
\left.
\begin{array}{lcll}
\lim_{a \to 0} \langle \rho(a,q,p) \rangle &=& 0 &\text{   if   } q<q^*  \ ,\\
&=& \frac{\Gamma(p-q^*/q)}{\Gamma(p)\Gamma(1-q^*/q)} & \text{   if   } q>q^* 
\end{array}
\right\}
\label{eq-expectetpart}
\end{equation}

The expected behavior recalled in Eq.~(\ref{eq-expectetpart}) is now tested numerically, in the context of CPM, by means of Monte-Carlo simulations.
The $\rho(a,q,p)$, as defined in Eq.~(\ref{eq-partratio}) above, are computed over $500$ independent realizations of CPM of length $2^{22}$, for $q =1,\ldots,15$, $p=2, 4, 5$ and $a=2^j$, with $j=1,\ldots,18$. 
The ensemble average $ \langle \rho(a,q,p) \rangle $ is estimated by the average $ \langle \rho(a,q,p) \rangle_{MC} $ of the $\rho(a,q,p) $ over the independent realizations.
The expected $ \langle \rho(a,q,p) \rangle $, according to Eq.~(\ref{eq-expectetpart}), and $ \langle \rho(a,q,p) \rangle_{MC} $ are compared in Fig.~\ref{figa}.
For $q \ll q^*$, $ \langle \rho(a,q,p) \rangle_{MC}  \simeq  \langle \rho(a,q,p) \rangle \equiv 0$, for all $p> 1$ and $a>0$.
For $q \gg q^*$, $ \langle \rho(a,q,p) \rangle_{MC}$ departs unambiguously from $0$,
for all $p> 1$ and $a>0$, and moreover follows a dependence in $q$ and $p$, that globally matches that of $ \langle \rho(a,q,p) \rangle $, expected from Eq.~(\ref{eq-expectetpart}).
The transition from zero to non-zero values of $ \langle \rho(a,q,p) \rangle_{MC}  $  occurs for values of $q$ typically around $q^*$, as theoretically computed from Eq.~(\ref{eq-qc}).

The match between the expected $ \langle \rho(a,q,p) \rangle $, according to Eq.~(\ref{eq-expectetpart}) and $ \langle \rho(a,q,p) \rangle_{MC} $ is not perfect though.
This may stem from a number of causes. On the one hand, the $\{ E_k, k=1,\ldots, n\}$ in the REM and the $\{ \ln T(a,t_k), k=1,\ldots,n_a \}$ in CPM, though both heavy-tailed might have not exactly the same distributions. On the other hand, the derivation of the theoretical results in Eq.~(\ref{eq-expectetpart}) relies on an exact independence assumption of the $\{ E_k, k=1,\ldots, n\}$, while the $\{ \ln T(a,k), k=1,\ldots,n_a \}$ still remain significantly correlated, as predicted by Eq.~(\ref{eq-cpcdep}}), which may affect the limiting ensemble average. 
Note that results are shown for the arbitrary scale $a=4$ only, as all conclusions drawn above are identical at all scales. 

These empirical observations are regarded as satisfactory results, corroborating numerically the theoretical analysis of the linearization effect observed in multifractal analysis and conducted from REM-type arguments. 

\begin{figure}[t]
\centerline{\includegraphics[width=70mm]{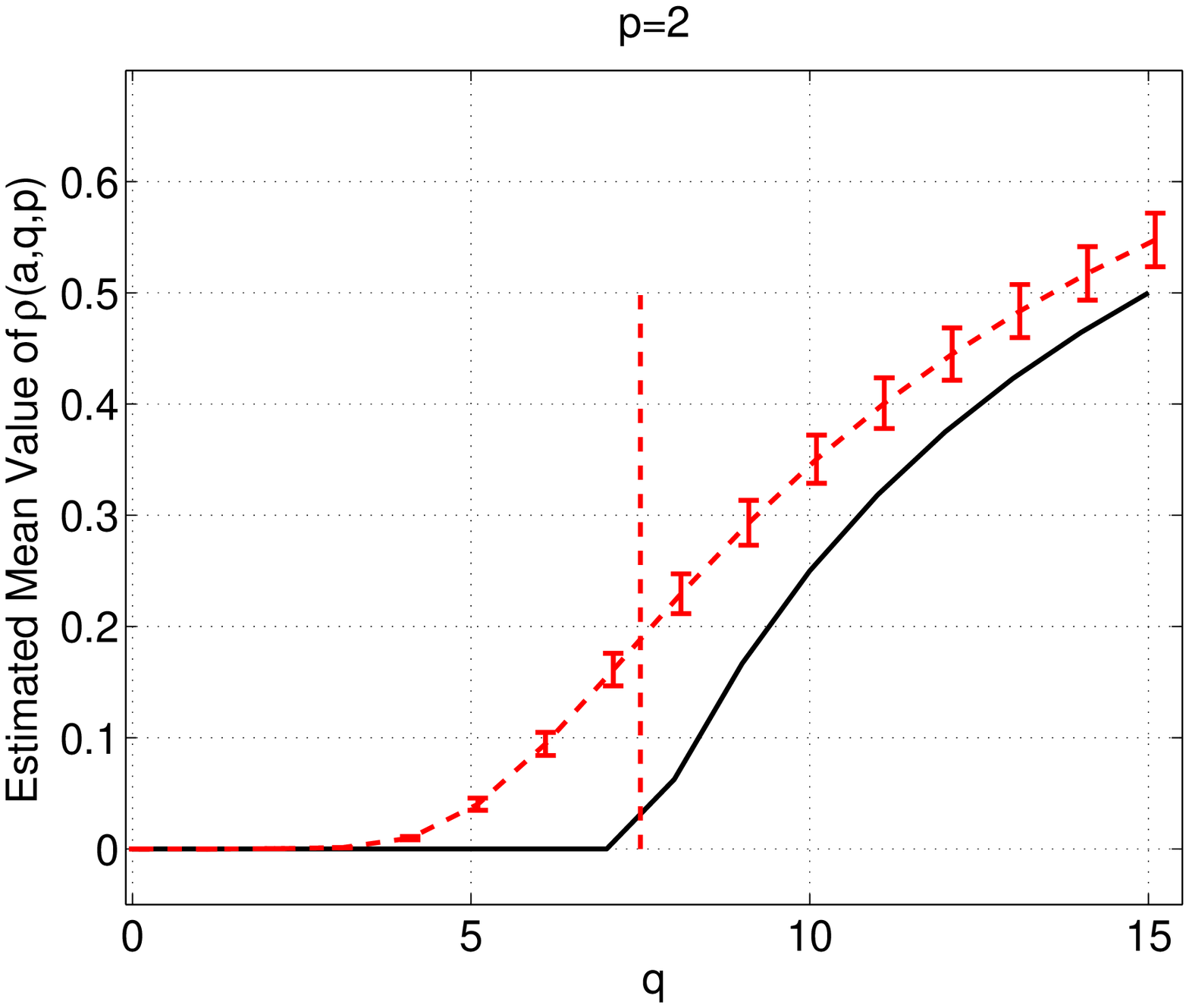}}
\centerline{\includegraphics[width=70mm]{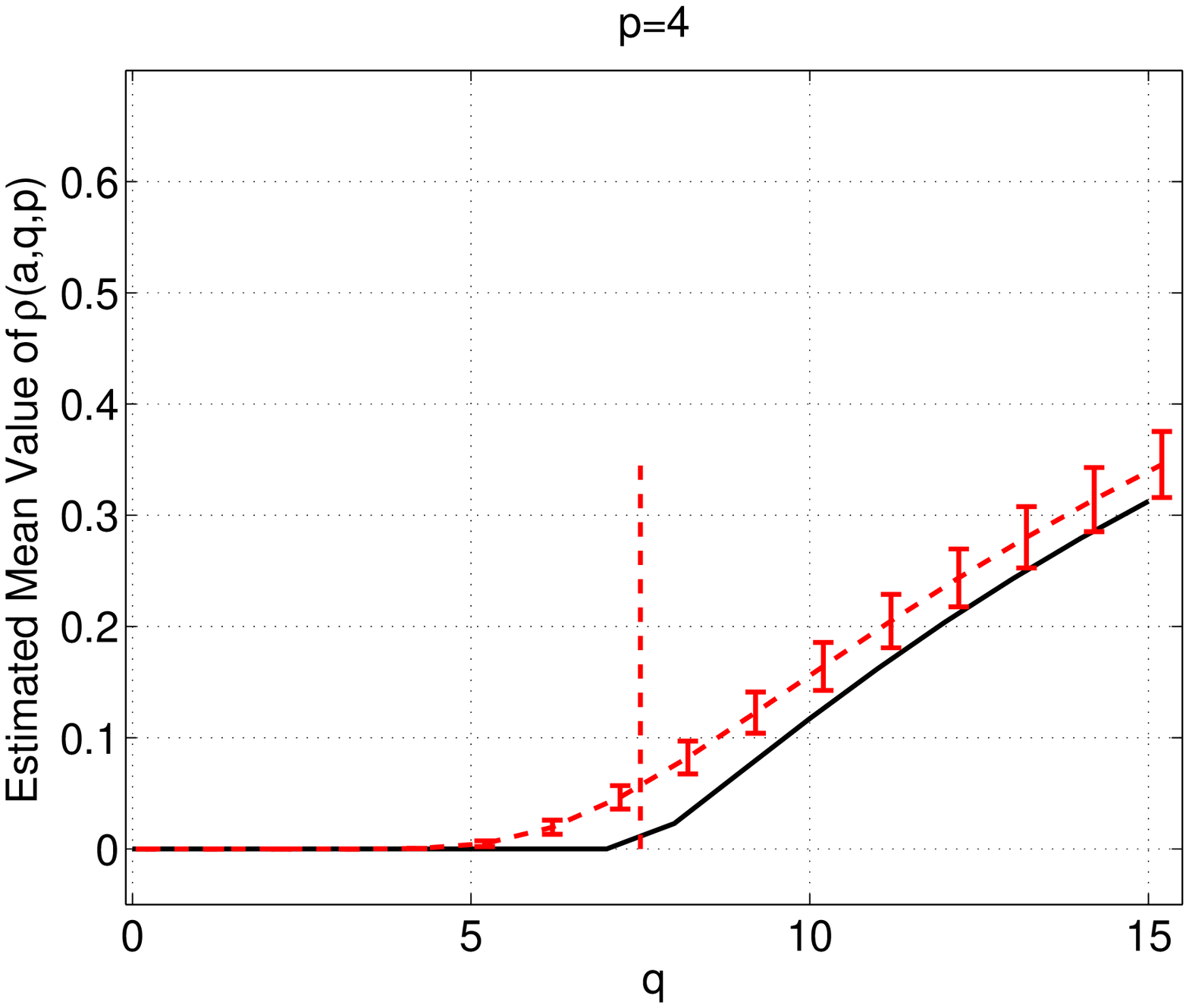}}
\centerline{\includegraphics[width=70mm]{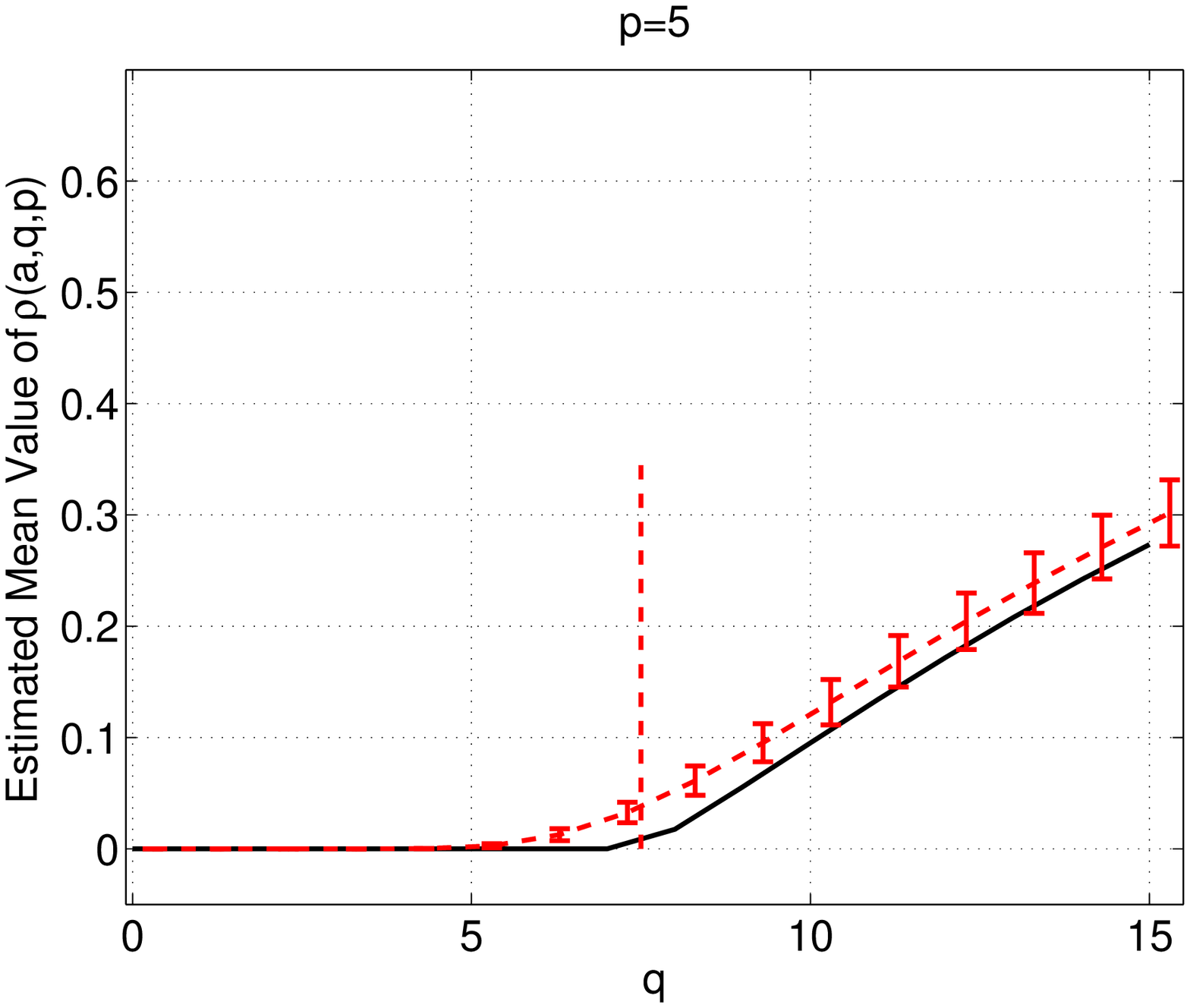}}
 \caption{\label{figa} {\bf Participation ratio.} Solid black line: $\langle \rho(a,q,p) \rangle $, according to Eq. \protect (\ref{eq-expectetpart}) ; dashed red line: $ \langle \rho(a,q,p) \rangle_{MC} $ averaged over Monte-Carlo simulations, with $95 \% $ confidence intervals ; red vertical dashed line: position of the critical $q^*$ as computed from Eq. \protect (\ref{eq-qc}). Top: $p=2$, middle $p=4$, bottom $p=5$, for scale $a=4$.} 
\end{figure}

\section{Discussion and conclusion} 

In this contribution, it has been shown that the time averages (or structure functions) of the ($q-$th power of the) increments of the sample path of Compound Poisson Motion, chosen as a simple representative of multifractal processes, cease to account correctly for the (ensemble average) moments of order $q$, above a critical order $q^*$.
This critical order is entirely defined from quantities entering the definition of the process and does not depend on the sample size of the observation: Increasing this sample size (by decreasing the sampling period $\delta t$) does not permit to increase the  range of orders $q$ for which moments can be correctly estimated. 
This critical order is not related either to the lack of finiteness of the moments. 
Moreover, for $q \geq q^*$, the structure function still exhibits power-law behaviors with respect to scale $a$, with scaling exponents that however behave linearly in $q$. 
Both the critical order $q^*$ and the slope of the linear behavior are predicted quantitatively.
These predictions are obtained from the tailoring of statistical physics arguments involved in the analysis of the REM to multifractal processes.
The reason why increasing the sample size does not permit a correct computation of the moments for $q>q^*$ can be understood as the \emph{non-self-averageness} property in the glass phase of the REM.
This correspondence is reminiscent of the analogy that leads from the thermodynamical formalism to the multifractal formalism, commonly used to measure the multifractal spectrum from empirical data \cite{fp85, arneodo1995}. 

This contribution can hence be read as a further effort to make explicit the fruitful correspondences between the thermodynamical and multifractal formalisms, in the spirit of e.g., \cite{fp85,PaladinVulpiani1987,arneodo1995,Frisch1995},
with a specific emphasis on  marrying in a single point of view different perspectives on the linearization effect: that of stochastic process sample path based statistical estimation, that of statistical physics and that of local regularity functional analysis.

Monte-Carlo simulations, based on the numerical synthesis of independent sample paths of CPM and estimation of the participation ratio, a classical tool in the statistical physics of condensed matter, satisfactorily confirm these predictions. 
These predictions based on REM-type statistical physics arguments are in perfect consistence with those proposed in \cite{lac04,ABRY:2007:B}, based on an extreme value analysis of the multifractal formalism and provides a complementary understanding of why time averages do not converge to ensemble averages.

The analysis conducted here can be, mutatis mutandis, applied straightforwardly to other multifractal processes. 
This is notably the case for fractional Brownian motion in multifractal time \cite{Mandelbrot1999}, which is obtained by subordinating CPM to a classical fractional Brownian motion and which constitutes a very appealing model to account for the multifractal properties of real data. 
Monte-Carlo simulations, not reported here, such as those described in Section~\ref{sec-mc}, performed on fractional Brownian motion in multifractal time, yield conclusions in perfect consistency with those drawn from the analysis of CPM. 

In addition, the analysis conducted here can also naturally be extended to multiresolution quantities other than the increments. 
We performed Monte-Carlo simulations on CPM and fractional Brownian motion in multifractal
time (not shown here) using increments of order $P$ (i.e., increments of increments of increments\ldots) as well as wavelet coefficients (computed from mother wavelets with different number of vanishing moments, see e.g., \cite{Mallat1998}). These simulations also lead to comparable
conclusions. 

Furthermore, this work opens the track for a systematic definition and estimation of a critical order for the moment estimation, in different contexts where the variables of interest consist of random exponentials, as in \cite{BenArous}.
This, together with the practical estimation of the critical order from a finite sample size observation, is under current investigation \cite{AngelettiBertinAbry2011}.

\section*{Acknowledgements}

PA and MM gratefully acknowledge the organizers of the 2008 edition of the Peyresq summer school in Signal and Image Processing, where this work was originally envisaged.


\appendix

\section{Random Energy Model}
\label{app-REM}

A very simple disordered model, which nevertheless captures a lot of the
phenomenology of realistic disordered systems, has been proposed
by Derrida, and called Random Energy Model (REM)
\cite{Derrida}.
It can be thought of as a spin model, although spins do not play any
essential role in the description. 
Considering a system of $N$ spins,
the corresponding number of configurations is $n=2^N$.
To each configuration $j$ is associated a random energy $E_j$
drawn at random from a distribution $P(E)$:
\be
P(E) = \frac{1}{\sqrt{N\pi J^2}} \exp\left(-\frac{E^2}{NJ^2}\right)
\ee
The energies $E_j$ are independent and identically distributed random variables.
We denote as $\rho(E) dE$ the number of configurations with energy in the
interval $[E,E+dE]$, so that $\rho(E)$ is the density of configurations with
energy $E$. The density $\rho(E)$ is a random quantity, but its fluctuations
are small if $\rho(E)$ is large, namely $\rho(E) \approx \langle \rho(E)\rangle$.
By definition, $\langle \rho(E)\rangle = n P(E)$, leading to
\bea \nonumber
\langle \rho(E)\rangle &=& \exp\left( N\ln 2 -\frac{E^2}{NJ^2}\right)\\
&=& \exp\left[\ln n \left(1 -\frac{\varepsilon^2}{J^2 \ln 2}\right)\right]
\eea
where the energy density $\varepsilon=E/N$ has been introduced.
One sees that if $1-\varepsilon^2/(J^2 \ln 2)>0$, corresponding to
$|\varepsilon|<\varepsilon^{\dag} = J\sqrt{\ln 2}$,
$\langle \rho(E)\rangle$ is exponentially large
with $N$, so that there is a large number of configurations at energy
density $\varepsilon$, and the assumption $\rho(E) \approx \langle \rho(E)\rangle$ is justified.
In contrast, if $|\varepsilon|>\varepsilon^{\dag}$,
$\langle \rho(E)\rangle$ is extremely small for large $n$. This means that in most
samples, there are no configurations at energy density $|\varepsilon|>\varepsilon^{\dag}$.
The non-zero, but small value of $\langle \rho(E)\rangle$ comes from the contribution
to the average value of very rare samples, which include some configurations
with exceptionally low (or high) energy.

We can now evaluate the partition function of the REM, defined as
\be
Z = \sum_{j=1}^{2^N} e^{-E_k/T}.
\ee
This partition function is a random variable, the typical value of which can be evaluated
as follows:
\be
Z \approx Z_{\mathrm{typ}}= \int_{-\varepsilon^{\dag}}^{\varepsilon^{\dag}}
d\varepsilon\, \langle\tilde{\rho}(\varepsilon)\rangle\,
e^{-N\varepsilon/T},
\ee
with the notation $\tilde{\rho}(\varepsilon)=N \rho(N\varepsilon)$. In the above equation, we have
replaced $\tilde{\rho}(\varepsilon)$ by $\langle \tilde{\rho}(\varepsilon)\rangle$
for $|\varepsilon|<\varepsilon^{\dag}$,
and by $0$ for $|\varepsilon|>\varepsilon^{\dag}$.
We can then write
\be
Z_{\mathrm{typ}} = \int_{-\varepsilon^{\dag}}^{\varepsilon^{\dag}} d\varepsilon\,
e^{-(\ln n)\, g(\varepsilon)}
\ee
with
\be
g(\varepsilon) = \frac{\varepsilon^2}{J^2 \ln 2}+\frac{\varepsilon}{T\ln 2} -1.
\ee
In the large $n$ limit, we can evaluate $Z_{\mathrm{typ}}$ through a saddle-point
calculation, namely
\be
Z_{\mathrm{typ}} \approx e^{-(\ln n) \, g_{\mathrm{min}}(\varepsilon^{\dag})}
\ee
where $g_{\mathrm{min}}(\varepsilon^{\dag})$ is the minimum value of $g(\varepsilon)$ over the interval
$[-\varepsilon^{\dag},\varepsilon^{\dag}]$.
Let us first consider the value $\varepsilon_{\mathrm{m}}$ which minimizes $g(\varepsilon)$
over the entire real line. Taking the derivative of $g(\varepsilon)$, one has
\be
g'(\varepsilon) = \frac{2\varepsilon}{J^2 \ln 2}+\frac{1}{T\ln 2}.
\ee
From $g'(\varepsilon)=0$, we find
\be
\varepsilon_m = -\frac{J^2}{2T}.
\ee
As $g(\varepsilon)$ is a parabola, it decreases for $\varepsilon<\varepsilon_{\mathrm{m}}$
and increases for $\varepsilon>\varepsilon_{\mathrm{m}}$.
If $\varepsilon_{\mathrm{m}}>-\varepsilon^{\dag}$, then $g_{\mathrm{min}}(\varepsilon^{\dag})=g(\varepsilon_{\mathrm{m}})$, so that
\be
Z_{\mathrm{typ}} \approx e^{-Ng(\varepsilon_{\mathrm{m}})}.
\ee
The condition $\varepsilon_{\mathrm{m}}>-\varepsilon^{\dag}$ translates into $T>T_{\mathrm{g}}$, where the
so-called glass transition temperature $T_{\mathrm{g}}$ is defined as
\be
T_{\mathrm{g}} = \frac{J}{2\sqrt{\ln 2}}.
\ee
For $\varepsilon_{\mathrm{m}} <-\varepsilon^{\dag}$, or equivalently $T<T_{\mathrm{g}}$, $g(\varepsilon)$ is an increasing
function of $\varepsilon$ over the entire interval $[-\varepsilon^{\dag},\varepsilon^{\dag}]$, so that
$g_{\mathrm{min}}(\varepsilon^{\dag})=g(-\varepsilon^{\dag})$, and
\be
Z_{\mathrm{typ}} \approx e^{-Ng(-\varepsilon^{\dag})}.
\ee
From these estimates of $Z_{\mathrm{typ}}$, one can compute the free energy
$F=-T\ln Z_{\mathrm{typ}}$, and the entropy $S=-\partial F/\partial T$.
For $T>T_{\mathrm{g}}$, one finds
\be
F=-N \left( T\ln 2+\frac{J^2}{4T} \right),
\ee
leading for the entropy to
\be \label{S-TgtTg}
S = N \left( \ln 2-\frac{J^2}{4T^2}\right).
\ee
For $T<T_{\mathrm{g}}$, we have
\be
F = TNg(-\varepsilon^{\dag}) = -NJ\sqrt{\ln 2}.
\ee
The free energy does not depend on temperature in this range, so that the
corresponding entropy vanishes:
\be
S=0,\qquad T<T_{\mathrm{g}}.
\ee
It can also be checked that the entropy given in Eq.~(\ref{S-TgtTg})
for $T>T_{\mathrm{g}}$ vanishes continuously for $T\to T_{\mathrm{g}}$.
Hence the temperature $T_{\mathrm{g}}$ corresponds to a glass transition temperature,
where the entropy goes to zero when lowering temperature, and remains
zero below $T_{\mathrm{g}}$. Actually, to make the statement sharper, only
the entropy density $S/N$ goes to zero for $T<T_{\mathrm{g}}$, in the infinite $N$
limit. Computing subleading corrections to the entropy $S$, one finds that
$S$ is independent of $N$, but non-zero, for $T<T_{\mathrm{g}}$.
The entropy is then intensive in this temperature range,
meaning that only a finite number of configurations,
among the $n=2^N$ possible configurations, are effectively occupied:
the system is quenched in the lowest energy configurations.

\section{Almost sure convergence of $\ln S/|\ln a|$}
\label{app-as-conv}

The aim of this appendix is to sketch the proof that for $n_k=2^k$ and $a_k=2^{-k}L$: 
\begin{equation} 
\label{app-eq-qed}
 \lim_{k\rightarrow +\infty} \frac{\ln S_{n_k}(a_k,q)} {|\ln a_k|}  \as = \lim_{a\rightarrow 0} \frac{\ln  M(a,q)}{|\ln a|}. 
\end{equation}
We refine the definition of $h^\dag$ by choosing
\begin{equation}
 \tau_k= \frac{1}{k^2}
 \end{equation}
in Eq.~(\ref{eq-hdag-na}), instead of a constant $\tau$.
This alteration does not affect $h_0^\dag$ due to the property
\begin{equation}\lim_{k\rightarrow +\infty} \frac{\ln \tau_k}{|\ln a_k|}=0.\end{equation}
With this choice of $\tau_k$, we have 
\begin{equation} \sum_k P(\exists i<n_k,\, h_{a_k}(i) < h^{\dag}_k) <+\infty.\end{equation}
The Borel-Cantelli lemma states that if a sequence of events $A_k$ satisfies $ \sum P(A_k) < +\infty$, then the event $A_k$ only happens for a finite number of $k$. Choosing the events to be $A_k=(\exists i<n_k,\, h_{a_k}(i) < h^{\dag}_k)$, we find that for sufficiently large $k$, all the $h_a$ are almost surely larger than $h^\dag_k$ . 
Denoting $\chi_I$ the characteristic function of the set $I$,
\begin{equation}
\Char{I}{x} = \begin{cases} 1 & \text{if } x\in I \\ 0 & \text{otherwise} \end{cases},
\end{equation}
we can write
\begin{equation} 
 S_{n_k}(a_k,q) \as{=} \frac{1}{n_k}\sum_{i=1}^{n_k} a^{q h_i} \Char{[h^{\dag}_k,+\infty)}{h_i}. 
 \end{equation}

In order to refine this result, we partition the interval $[h^\dag,+\infty)$ in different sub-intervals.
First, we define a separation point $c_{\infty}$ by  
\begin{equation} \psi'(c_{\infty})=0.\end{equation}
The corresponding sub-interval is
\begin{equation} I_{\infty}=(c_{\infty},+\infty). \end{equation}
The remaining interval $[h^\dag_k,c_{\infty}]$ has a finite length $l_k$ 
\begin{equation} l_k=c_{\infty}-h^\dag_k. \end{equation}
We partition this interval into a number $[2\ln n_k]$ of sub-intervals
(where $[x]$ denotes the integer part of $x$), with
\begin{equation} c_p=h^\dag+ p\frac{l_k}{1+[\ln n_k]} \end{equation}
\begin{equation} I_p = [c_p ,  c_{p+1} ], \quad |p|< 1+[\ln n_k] \end{equation}

We call then $m_p$ the density of points inside the interval $I_p$
\begin{equation}
 m_p = \frac{1}{n_k} \sum_{i=1}^{n_k}   \Char{I_p}{h(i)}.
\end{equation}
The quantity $m_p$ can be bounded using the Borel-Cantelli lemma, leading to 
\begin{equation}
  \frac{\langle I_p \rangle}{2} \as{<} m_p \as{<} (\ln n_k)^3 \langle I_p \rangle
\end{equation}

The upper bound is a direct consequence of the classical Markov's inequality. 
However the lower bound is more subtle because it requires the use of the Chebyschev 's inequality and consequently a bound for the correlation of the $\chi_{I_p}$. Let us make the realistic assumption that such a bound exists.
Moreover the vanishing length of the intervals $I_p$ implies that :
\begin{equation} \langle I_p \rangle \underset{k\rightarrow +\infty}{\sim} \frac{l_k}{\ln n_k} a^{\psi(c_p)} \end{equation}
Hence, we have $\varepsilon_1\in(0,1)$ such that for sufficiently large $k$,
\begin{equation}
S_{n_k}(a_k,q) \as{<} \sum_{p} (\ln n_k)^2 l_k (1+\varepsilon_1) a^{q c_p + \psi(c_p)} + m_{\infty} a^{c_{\infty}}.
\end{equation}
We are mainly interested in the extremal contribution, which comes from $c_m$ : 
\begin{equation}
(1-\varepsilon_{2}) l_k \frac{a^{q c_m +\psi(c_m)} }{2 \ln k}   \as{<} S_{n_k}(a_k,q) \as{<} (1+\varepsilon_{3}) (\ln n_k)^3  l_k  a^{q c_m +\psi(c_m), }
\end{equation}
with $\varepsilon_2,\varepsilon_3 \in (0,1)$.
These two inequalities can be rewritten in logarithmic terms,
\begin{equation}
\label{app-eq-as}
\begin{split}
 -(q c_m +\psi(c_m)) &+\frac{\ln(1-\varepsilon_{2})+\ln l_k -\ln 2-\ln \ln n_k }{|\ln a|}\\ &\as{<} 
\frac{\ln S_{n_k}(a_k,q)} {|\ln a_k|} \as{<} \\ -(q c_m +\psi(c_m)) &+\frac{3 \ln \ln n_k+\ln(1+\varepsilon_{3})+\ln l_k }{|\ln a|}. 
\end{split}
\end{equation}
The total length $l_k$ is bounded because so are $h^\dag_k$ and $c_{\infty}$.
When $k\rightarrow+\infty$, the length of the interval $I_p$ tends to $0$, which means that
\begin{equation}
c_m \rightarrow \max(h_m,h^\dag),
\end{equation}
Thus taking the limit $k\rightarrow +\infty$ in Eq.~(\ref{app-eq-as}) leads to
\begin{equation}
\lim_{k\rightarrow +\infty} \frac{\ln S_{n_k}(a_k,q)} {|\ln a_k|} \as{=} - q \max(h_m,h^\dag) - \psi (\max(h_m,h^\dag)).
\end{equation}
The right hand side of the previous equation exactly corresponds to the evaluation of the truncated moment by the saddle-point method obtained in Eqs.~(\ref{eq-momHbis}) and (\ref{eq-momHter}), which demonstrates the validity of Eq.~(\ref{app-eq-qed}).
A similar result for the dyadic multiplicative cascade (cf \cite{Ossiander2000}) suggests that the almost sure convergence holds in the general case even without the assumption made in order to obtain the lower bound.

 \bibliographystyle{plain}
\bibliography{remmf}

\end{document}